# The Design and Operation of Rules of Origin in Greater Arab Free Trade Area: Challenges of Implementation and Reform

Bashar H. Malkawi* and Mohammad I. El-Shafie**

**Abstract**

Rules of origin (ROO) are pivotal element of the Greater Arab Free Trade Area (GAFTA). ROO are basically established to ensure that only eligible products receive preferential tariff treatment. Taking into consideration the profound implications of ROO for enhancing trade flows and facilitating the success of regional integration, this article sheds light on the way that ROO in GAFTA are designed and implemented. Moreover, the article examines the extent to which ROO still represents an obstacle to the full implementation of GAFTA. In addition, the article provides ways to overcome the most important shortcomings of ROO text in the agreement and ultimately offering possible solutions to those issues.

## I. Introduction

Rules of Origin (ROO) mechanism is used to determine the origin of a product. ROO serve many purposes such as collecting data on trade flows, implementing preferential tariff treatment, and applying anti-dumping duties.[1] ROO can be divided into preferential and non-preferential rules. Preferential ROO are used to determine whether a product originates in a preference-receiving country or trading area and hence qualifies to enter the importing country on better terms than products from the rest of the world.[2] Non-preferential ROO are used for all other purposes, including enforcement of product- and country-specific trade restrictions that increase the cost of, or restrict or prevent, market entry. Preferential ROO differ from non-preferential ones because they are designed to minimize trade deflection.[3]

The key challenges of constructing ROO in preferential trading relationships are first, finding the balance between the effectiveness and the efficiency of ROO, and second, simplifying them and making them more transparent.[4] In principle, ROO are supposed to be straightforward and easy-to-follow methods used to determine origin especially when a product is manufactured in one country, which rarely happens in reality. However, more than often, ROO are complex and protectionist method used a barrier to trade. The higher the level of processing that is required in the partner the more difficult it will be for the partner's products to receive trade preferences, especially if the partner is a small low-income, low-wage economy. Further, restrictive rules of origin not only affect trade within a region but also the ability of local firms to participate in global trade networks. Thus, the nature of the rules of origin can act to undermine the stated intentions of preferential trade agreements to stimulate regional trade but also to compromise broader trade objectives[5].

---

* Bashar H. Malkawi is Dean and Professor of Law at the University of Sharjah, United Arab Emirates. He holds S.J.D in International Trade Law from American University, Washington College of Law and LL.M in International Trade Law from the University of Arizona.
** Mohamed I. El-shafie is Associate Professor of Law at the University of Sharjah. He holds Ph.D from University of Paris 1 (la Sorbonne), France.
This research was funded by grant from the University of Sharjah.

[1] See Edwin Vermulst, Rules of Origin as Commercial Policy Instruments - Revisited, 26(6) Journal of World Trade 61, 63 (1992). See also Maarja Saluste, Rules of Origin and the Anti-Dumping Agreement, 12 Global Trade and Customs Journal 54, 57 (2017).
[2] See Joseph A. LaNasa III, Rules of Origin and the Uruguay Round's Effectiveness in Harmonizing and Regulating them, 90 A.J.I.L. 625, 627 (1996).
[3] Trade deflection occurs when a company undertakes minimal processing or assembly in a preference-receiving country to take advantage of preferences. See Kathryn L. McCall, What is Asia Afraid Of? The Diversionary Effect of NAFTA's Rules of Origin on Trade Between the United States and Asia, 25 Cal. W. Int'l L.J. 389, 393 (1995). See also Ann Van de Heetkamp and Ruud Tusveld, Why Is this Complex? Origin Management: Rules of Origin in Free Trade Agreements 183 (2011).

While ROO intended to facilitate trade among Arab countries and keep free riders from profiting unwarrantedly from Greater Arab Free Trade Area (GAFTA), ROO should not become instruments of forced investment and cumbersome for traders so that they forgo using GAFTA preferences.[6] ROO GAFTA deflects non-partner country goods from profiting from the favorable tariff rates. GAFTA ROO determines the tariff levied on each product imported into an Arab country.

The cornerstone of GAFTA is the preferential tariffs for GAFTA-origin products. GAFTA eliminate all tariffs on products that originate in participating countries within no more than ten years.[7] In most instances, higher "most favored nation" tariffs remain applicable to imports from outside the region.[8]

The article begins in section I with a brief introduction on the functioning of ROO. Section II analyzes the different ROO and common provisions used in GAFTA accompanied by examples, when appropriate, to illustrate the rules. Section III analyzes other relevant issues for GAFTA ROO such as operations not conferring origin and cumulation. Section IV addresses the issue of reviewing Customs Decisions regarding GAFTA ROO. Section V analyzes the status of adopting ROO in Arab countries and the gaps that exist in relation to implementation. Section VI describes certificate origin as currently implemented in GAFTA and its regulations. Section VII provides statistics as to inter-trade among Arab countries and measures the possible impact of ROO on trade among trading partners. Section VIII concludes with a set of conclusions and recommendations on how ROO should be reformed and considered in order to facilitate trade among GAFTA countries.

---

[4] See Vivian C. Jones and Michael F. Martin, International Trade: Rules of Origin, Congressional Research Service, CRS Report for Congress, p. 7 (2012).
[5] See Paul Brenton et al, Rules of Origin and SADC: The Case for Change in the Mid Term Review of the Trade Protocol, Africa Region Working Paper Series, 83 (June 2005).
[6] See Ahmed Al-Kawaz, Occupational Restructuring of Non-oil Manufacturing Labor Force: The Case of Kuwait 8-10 (Arab Planning Institute, 2010).
[7] See GAFTA, Part 2, art. 2.1, Executive Program for Implementing GAFTA. This does not apply to all products such as agricultural products.
[8] Arab countries maintain different MFN applied tariff rate: Algeria 18.8, Bahrain 4.7, Djibouti 20.9, Egypt 16.8, Jordan 10.2, Kuwait 4.7, Lebanon 5.7, Mauritania 12.0, Morocco 11.2, Oman 4.7, Qatar 4.7, Saudi Arabia 5.1, Sudan 21.2, Syria 16.5, Tunisia 14.1, UAE 4.7, Yemen 7.5. See WTO, ITC, and UNCTAD, World Tariff Profiles 6-10 (2015).

**II. Rules of Origin in GAFTA**

The most important attempts to achieve economic integration among Arab countries were the Common Market of 1964, and the Agreement of 1981 on the Facilitation and Development of Trade. During their summit in Cairo, Egypt in 1996, Arab countries took the decision to revive the 1981 agreements and create GAFTA.[9] The agreement of GAFTA entailed, among other things, the elimination of non-tariff barriers and the reduction of tariff rates on goods.

The Economic and Social Council of the League of Arab States (LAS) declared an executive program for GAFTA.[10] According to that decision, it was expected that the implementation of the executive program would be accomplished on 21/12/2007. However, in reality, the first steps for implementing the executive program have not taken place until the year 2005.

GAFTA includes general rules of origin comprising 13 pages.[11] In addition to the general rules of origin, GAFTA and implementing regulations and decisions include detailed ROO, which covers approximately 100 pages. GAFTA implementing regulations and decisions provide in exhaustive detailed and specific ROO for each product which determines the process required of exporters and importers in order to qualify for preferential tariff benefits under GAFTA.

Without the inclusion of these ROO, it would allow some Arab countries, especially the ones that enjoy low labor cost, to serve as an export platform, where foreign companies could establish final assembly facilities which can import parts and components from Asia such as China, and then assemble them in that particular Arab country. GAFTA desires to encourage manufacturers in Arab countries to go beyond serving as staging grounds for final assembly and into the manufacture of major parts and components anywhere in the region. Of course, if there were a customs union among Arab countries ROO would become irrelevant since all imported goods would face the same tariff level thus it does not matter from where these imported goods would enter the region.[12] However, as long as there are differences in the tariffs from one Arab country to another, then the preferred place of entry will be the one that provides for the lower tariff, from which goods then can be circulated throughout the region.[13]

GAFTA formulated ROO that have four characteristics: consistency, uniformity, objectivity, and reasonableness.[14] However, achieving such objectives is important, but unfortunately this has not been always the case given the complicated nature of GAFTA general ROO and the other varied rules.

---

[9] See Tamer Mohamed Ahmed Afifi, The Challenge of Implementing the Overlapping Regional Trade Agreements in Egypt 19-20 (2007).
[10] See Economic and Social Council of the League of Arab States, Decision no. 1317/59 (March 19, 1997).
[11] See GAFTA, Part 4, Executive Program for Implementing GAFTA.
[12] See Daniel Schwanen, A NAFTA Customs Union: Necessary Step or Distraction? 60 International Journal 399, 400 (2005) (the need for verifying the origin of goods circulating across borders within the customs union disappears. A customs union would thus reduce the need for border checks on trade and remove other important trade distortions within the NAFTA).
[13] See WTO, ITC, and UNCTAD, World Tariff Profiles, *supra* note 6.
[14] See Rule 11, Chapter 4, GAFTA. Article 2(e) of the WTO Agreement on Rules of Origin provides that WTO member countries must ensure that "their rules of origin are administered in a consistent, uniform, impartial and reasonable manner." See also John J. Barceló III, Harmonizing Preferential Rules of Origin in the WTO System, Cornell Law Faculty Publications Paper No. 72, p. 7 (2006).

GAFTA adopted several general criteria to determine origin of goods. These rules are:

**1) Goods that are wholly produced in one country**. Under the "wholly obtained or produced" rule of origin, in order for a good to qualify for preferential treatment, the product must be "wholly" the growth, production or manufacture of a GAFTA member state. Also, the concept of "wholly" should be interpreted narrowly since all the inputs must be produced in the exporting country to qualify for preferential treatment; third party inputs are not allowed. GAFTA provides a list of goods to be considered "wholly produced". The list covers primary products, raw minerals, lumber, and unprocessed agricultural commodities. Examples of these goods include live animals born and raised in an Arab country (chapter 01), live trees and other plants harvested or gathered in a single country (chapter 06), edible fruit and nuts; peel of citrus fruit or melons (chapter 08), cereals (chapter 10).

Fisheries can qualify for preferential treatment if they caught in a country's river (chapter 03). For fisheries captured in high seas or outside of territorial waters, they will originate in GAFTA if the vessel is owned by the beneficiary country. This means that flag, registration, and ownership will determine the nationality of the vessel and thus meeting GAFT rules of origin. It is not obvious if there is a national officer and crew requirement. Rules of origin for fisheries caught in high seas ought to be relaxed by requiring nationality of vessels based on flag, registration, and ownership only.

The "wholly obtained or produced" rule of origin is straightforward since it does not include materials from third countries and thus does not pose problems.[15] The more difficult and more common situation occurs when particular parts of a product, often in the form of raw materials or components, come from more than one country.[16]

**2) Value-added content allows for imported goods** to receive preferential treatment under GAFTA only if no less than 40% percent of the total product's value is attributable to production in another Arab country using: "net cost" method, or the "final value of product" method. Examples of goods subject to this rule of origin include butter (sub-chapter 0405), non-organic chemical products (chapter 28), fertilizers (chapter 31), and electrical generators (sub-chapter 8504).

---

[15] Producers even of basic goods or raw materials are plagued with the various phases of a good's production to determine originating status. If a fish is caught, transported to shore, and then cleaned and filleted, the question arises if there has been a significant transformation of the original product. See Stefano Inama, The Uruguay Round Agreement on Rules of Origin, in Rules of Origin in International Trade 104 (2009).

[16] With the increasing trend that has been labeled as the "global factory," most final products in contemporary international commerce involve factors of production from more than one country. See Jacques HJ. Bourgeois, *Rules of Origin: An Introduction, in* RULES OF ORIGIN IN INTERNATIONAL TRADE **1,** 4-5 (Edwin Vermulst, Paul Waer & Jacques Bourgeois eds.,1994) (discussing this trend).

The "net cost" method aggregates the costs of the various inputs.[17] On the other hand, the "final value of product" method examines value of the finished products.[81] Under both methods, the non-regional materials (imported inputs) are subtracted.[18]

GAFTA uses ex-works price to calculate value-added.[19] The ex-works price exceeds ex-works cost and net cost since it includes additional factors such as manufacturing profit, packing cost, packing material, and royalty.

The "net cost" method requires accurate accounting for input, direct and indirect labor and overhead costs, and other costs.[20] Certain costs be ultimately included or excluded from the calculations. Parts and materials accounting can be straightforward. However, other costs, such as labor and overhead items which normally must be allocated over different product lines.[21] GAFTA does not provide details as how to allocate these costs. Indeed, this is where inconsistencies and uncertainties can arise. Costs may be altered to benefit a product line which is struggling to meet a 40% percent value-added when the other product line is either well above or well below the required threshold. The accounting can be complex. For example, if an Egyptian factory produces both pens and pencils in equal quantities. The materials cost for each is Egyptian Pound 100; however, labor on a per unit basis for the pen is Egyptian Pound 25, while it is Egyptian Pound 50 for the pencils. The issues that arises is whether labor charges such as supervisory personnel and factory overhead should be allocated equally to pens and pencils based on materials costs, or one third to pens and two thirds to pencils based on labor costs.

The "final value of product" method is generally simpler for manufacturers, importers and customs officials than calculating "net cost". With the" final value of product", only the value of the non-originating imported materials must be calculated.[22] The

---

[17] The added-value content of a product, where calculated on the basis of the "net cost " method shall be determined as follows:

   The value content as a percentage = value-added inputs/final value of the product X 100

See Rule 3, Chapter 4, GAFTA.

[81] The value content of a product, where calculated on the basis of the "final value" method, shall be determined as follows:

   The value content expressed as a percentage = (final value-value of non-originating materials)/final value of the product X 100

*Id.* See Rule 3, Chapter 4, GAFTA.

[18] See Catherine Truel, A Short Guide to Customs Risk 85-89 (2017) (The value added content rule demands the identification of both the value of the imported materials and the value produced in the country).

[19] Ex-works price means the price paid to the manufacturer in whose facility the last processing is carried out, provided the price includes the value of all the products used in the manufacturing minus any internal taxes which are, or may be, repaid when the product obtained is exported. See Antoni Estevadeordal and Kati Suominen, Rules of Origin in the World Trading System 61 (2003), available at < https://www.wto.org/english/tratop_e/region_e/sem_nov03_e/estevadeordal_paper_e.pdf> (last visited June 11, 2018).

[20] See Rule 3, Chapter 4, GAFTA.
[21] Overhead costs include charges for testing, research and development, insurance, royalty for intellectual property rights, and factory rental. *Id.*
[22] See David A. Gantz, A Post-Uruguay Round Introduction to International Trade Law in the United States, 12 Ariz. J. Int'l & Comp. Law 7, 140 (1995).

invoice prices of the components can be representative of the cost for the non-originating imported inputs. Thus, a production cost analysis with all its complexities is normally not required. The "final value of product" method can be difficult when there are many inputs from many different sources so it is hard to record-keep all the information, or if there is sourcing between related companies. For example, if an Egyptian producer of XYZ sells all of the products it produces to a related Lebanese importer. Here, for purposes of determining the final value, parties are related to each other can it can be difficult to do the accounting for the value of non-originating materials can be based.

GAFTA does not determine the situations whereby any of these two methods can be used. It is safe to say that whenever the use of either method is permitted, the exporter may use the most favorable method. For instance, suppose that the "net cost" of XYZ is 100. The "final value of product" is 120 and the value of non-originating materials imported from outside the region is 60. Under the "final value of product" method, value content is 50% percent [value content = (120-60)/120 X 100 = 50]. Under the "net cost" method, the value content is 40% percent [since value content = (100-60)/100 X 100 = 40]. In this typical example, the "final value of product" method produces big value content.

Customs authorities worldwide, including in Arab countries, face difficulties in verifying the value-added of inputs in order to determine preferential treatment.[23] GAFTA authorities may need to do away with the "net cost" method. If this seems not plausible, then the "final value of product" should be used in most cases. Exceptions for applying the "net cost" can be used in very limited cases such as when the "final value of product" cannot be used because of dealings between two related companies.

**3) Specific criteria such as a change of tariff category test using the Harmonized System (HS) nomenclature.**[24] Under this "change of tariff category" rule, when a change in tariff category occurs, the country where that shift occurs is the country of origin.[25] This most commonly occurs when a manufacturing process results in a product that is assigned a tariff heading different from the tariff heading of input or materials from outside the region. In other words, a "change in tariff category" rule requires classification of the good in question twice: once on arrival into an Arab country and again on departure from that country. The GAFTA "change of tariff category" rule is not completely different from the "substantial transformation" rule adopted by some countries like the U.S.[26] For instance, a UAE producer of matches imports a paperboard sheet from Morocco which is classified under harmonized tariff schedule subheading 4823.20. After importation, the producer cuts and folds the sheet and prints upon it the name of "al-ʿūd", and then inserts a UAE produced slide drawer with UAE produced matches into the sleeve. The finished matchbox is classified under harmonized tariff schedule subheading 4819.60. The change in tariff category rule is applied at the four-digit heading level. The imported paperboard sheets of heading 4823 undergo the specified change in tariff category to heading 4819, and the origin of the matchbox is UAE.

---

[23] See World Customs Organization, World Trends in Preferential Origin Certification and Verification, WCO Research Paper No. 20, 18 (November 2011)

[24] The Harmonized System has twenty-two sections divided into ninety-seven chapters and contains over 5,000 article descriptions using a six-digit description for all products. The Harmonized System headings are designed to progress from crude products to those based on increasingly sophisticated processing. The first two digits are the "Chapter" in which the product is contained. There are ninety-seven Chapters in the Harmonized System, reflecting the diversity of possible product categories. The first four digits taken together are called the "Heading" and provide a more specific description of the product. The last two digits of the six digits provide a still more specific level of description. See Harmonized Commodity Description and Coding System, United Nations International Trade Statistics Knowledge-base http://unstats.un.org/unsd/tradekb/Knowledgebase/Harmonized-Commodity-Description-and-Coding-Systems-HS (last visited Nov. 18, 2017).

[25] See Joseph A. LaNasa II, *Rules of Origin and the Uruguay Round's Effectiveness in Harmonizing and Regulating Them,* 90 AM. J. INT'L L. 625, 629-36 (1996) (all discussing this test in detail).

In GAFTA, there are about 112 HS subheadings that incorporate change of tariff category-based rules. This is a reminder to the importance of HS classification in determining origin. Examples of goods subject to GAFTA "change in tariff category" rule include furniture (chapter 94), foodstuff (chapter 21), raw minerals (chapter 26), natural and artificial fur (chapter 43), and silk (chapter 50).

It should be noted that GAFTA tariff change rule is strict since it requires – in most circumstances- that *all* of the parts used in manufacturing a product be imported and classified under a different tariff category than the category in which the final product would be classified. In order to facilitate trade, GAFTA "change of tariff category" rule should be relaxed. For example, there could a *de minimis* exception, in which parts comprising up to a certain percentage can be classified under the same tariff category of the final product.[27]

Not all tariff heading changes are considered significant enough to produce a change in origin. Use of tariff heading in GAFTA should necessitate the drafting of additional specifications to indicate which changes in tariff classification must occur to change the origin of imported materials.[28] This determination requires reviewing the tariff list product-by-product. Moreover, classification of goods is not always a simple exercise.[29] When the classification of a good and its materials is not known, making that determination can be a significant problem. Some Arab countries who export only a limited number of products could opt for an origin regime based on an across-the-board change in tariff heading.

---

[26] Substantial transformation means fundamental change in form, appearance, nature "or" character of article which adds to value of article. The U.S. Supreme Court defined substantial transformation further in Anheuser-Busch Brewing Association case. See Anheuser-Busch Brewing Assn. v. U.S., 207 U.S. 556, 561 (1908) (The court decided that manufacture implies a change, but every change is not manufacture, and yet every change in an article is the result of treatment, labor, and manipulation. There must be transformation: a new and different article must emerge, having a distinctive name, character, or use). Change in name only could not be in and of itself the determinative factor to meet the "substantial transformation" test. The CIT decided that the name factor in meeting the "substantial transformation" test is the weakest evidence of meeting the test. See Juice Prod. Assn. V. U.S., 628 F. Supp. 978, 989 (Ct. Intl. Trade 1986).

[27] In some parts, GAFTA adopts this approach by allowing up to 20 percent of *any* material to be used and that are classified under the same tariff category of the final product. See Economic Committee, Explanatory Notes: Item 6, List 1, Items whose ROO are Agreed Upon according to Change in Tariff Category Rule, Doc. No. G03-96/ (15/08)/05-m (0326)).

[28] HS codes are continuously amended; producers must constantly re-evaluate sourcing and production patterns, as fluctuations in HS codes trigger fluctuations in prices of production materials and associated costs. Stefano Inama, Drafting Preferential Rules of Origin, in Rules of Origin in International Trade (2009) 426.

[29] See David Palmeter, The WTO as a Legal System—Essays on International Trade Law and Policy 141; 150-152 (2003) (Although tariff heading changes method, though perhaps on balance the least costly and clearest, is nevertheless not without costs such as maintaining records of input origins, classification conundrums, and other disadvantages).

GAFTA could ease the burden of ROO by exempting certain goods from meeting the required ROO. In other words, in some cases and for some goods there ought not to be value-added, changes of tariff category, or certain process requirements. Simply, these goods can pass through some Arab countries. The exemption can apply to imported goods that have no tariffs at all or where trivial tariff is imposed.

**4) Certain process requirements for some products** allowing certain operations to confer origin. Goods that are subject to this rule of origin include soya-bean oil, ground-nut oil, olive oil, palm oil, sunflower-seed, safflower or cotton-seed oil, coconut, palm kernel or babassu oil, rape, colza or mustard oil, animal or vegetable fats and oils, margarine; edible mixtures or preparations of animal or vegetable fats or oils, and their fractions (sub-chapters 1507-1515), and goods listed in sub-chapters 2840, 4302, 5007, 5604. Although certain process requirements are intended to clarify the value-added determination for particular products, they are often complex.

**5) Combination of one or more of the above ROO.** For example, GAFTA requires some products satisfy the change in tariff heading test as well as contain a minimum level of domestic added value. Examples include organic chemicals (chapter 29), photographic or cinematographic goods (chapter 37), felt, whether or not impregnated, coated, covered or laminated (sub-chapter 5602). Based on the current text of GAFTA and its rules, there is no hierarchy in determining what rule of origin comes first when combining more than rule of origin.

GAFTA specified production methods for textiles and clothing products.[30] Although GAFTA claims duty free access for textile and clothing products, the specific ROO for these products are drafted to mitigate the likely effects of textiles and apparels trading on domestic clothing industry in Arab countries especially Arab producing countries such as Egypt, Morocco, and Tunisia.[31] For some intermediate products - such as woven fabrics of silk or of silk waste, textile wall coverings, and textile fabrics otherwise impregnated, coated or covered (sub chapter 5007)- GAFTA uses the "three operations" rule. Under the "three operations" rule, a textile product will be considered a product of an Arab country, if the fabric is *printed* in an Arab country and *the printing is accompanied by two or more of the following operations*: bleaching, shrinking, fulling, napping, decating, permanent stiffening, weighting, permanent embossing or moireing.[32] In other words, combination of processing operations –such as printing, shrinking, and weaving - confers origin. Under the "three operations" rule, it does not matter if for instance a fabric of silk or of silk waste is actually woven in an Arab country. The fabric of silk or of silk waste could be woven in Lebanon, but the printing, bleaching, and shrinking can occur in Jordan. For some other textile products – such as textile fabrics impregnated, coated, covered or laminated with plastics (sub chapter 5903)-, the "three operations" rule is used provided that the value of the unprinted fabric used does not exceed 47.5 per cent of the ex-work price of the product. This is a special technical requirement combined with a value added rule.

---

[30] Textile and clothing products are classified under chapters 50 – 63. Division can be made between textile products in chapters 50 – 60 and clothing products (finished products) in chapters 61 – 63.
[31] The reasons for protectionist policies in the textile sector are to be found in the importance of the textile sector for employment policy in some Arab countries. Textile is a labor-intensive industry which requires low skilled workers who if laid off could encounter hard time to find a new job. See Masakazu Someya, Hazem Shunnar and T.G. Srinvasan, Textile and Clothing Exports in MENA: Past Performance, Prospects and Policy Issues in Post MFA Context 6-8 (2002).
[32] An article usually will not be considered to be a product of a particular country by virtue of merely having undergone printing of fabrics or yarns (marginal operations). The printing confer origin when accompanied by two or more of the following finishing operations: bleaching, shrinking, fulling, napping, decating, permanent stiffening, weighting, permanent embossing, or moireing.

For some textiles and apparels (such as felt, textile fabrics, floor coverings, textile wicks, and headgear and parts thereof), they must be produced from yarn or fiber produced in the partner country.[33] This means that everything from the yarn forward up the production chain of an article must be of an Arab country origin. This may restrict the ability of a manufacturer to source its inputs. Furthermore, this rule may cause tariff escalation since the cost of using foreign yarn from a non-Arab country results in a higher tariff for the entire product. In a temporary fashion, GAFTA should allow for importation of apparel containing third-party content.

### III. Miscellaneous Provisions

GAFTA provides for operations that result from non-qualifying operations such as simple packaging operations, dilution with water, or any process. These operations may be used to circumvent GAFTA's ROO. Therefore, these kinds of operations do not confer origin.[34] For example, processing such as peeling, freezing, or cooking, may not significantly alters the good's character by changing its shape or size or its quality. To wit, such processing is not sophisticated and adds little value to the good in question. A combination of two or more insufficient processing operations does not confer origin, even if the product specific rules of origin have been satisfied. However, all operations carried out by the producer on a given product shall be considered together when determining whether the combined operations are to be regarded as insufficient.[35]

In cumulation, GAFTA allows for forty percent of Arab country input into another Arab country's goods, and vice versa, without conferring upon those goods the status of non-originating products.[36] In other words, GAFTA permits Arab countries to use inputs from a specific GAFTA member or group of GAFTA members without affecting the origin of the goods. For full cumulation, which allows the parties to an agreement to carry out working or processing on non-originating products in the area formed by them, Arab countries can examine the possibility of adopting full cumulation or even cross cumulation in the future.[37]

---

[33] Manufacture from fabric is not sufficient to confer origin on apparel. There is the requirement of double transformation i.e. from yarn to fabric and from fabric to clothing. See Jaime de Melo, Improving Market Access for Jordanian Exports to Europe, Economic Research Forum, available at < https://theforum.erf.org.eg/2018/03/19/improving-market-access-jordanian-exports-europe/> (last visited Oct. 24, 2018).
[34] See General and Detailed Principles of GAFTA Rules of Origin, art. 6.1, Doc. No. G03-19/13 (07/09)/02-t (0542)).
[35] *Id*. art. 6.2.
[36] See Rule 5, Chapter 4, GAFTA.
[37] See General and Detailed Principles of GAFTA Rules of Origin, art. 3, Doc. No. G03-19/13 (07/09)/02-t (0542)). See also Pamela Bombarda and Elisa Gamberoni, Firm heterogeneity, Rules of Origin and Rules of Cumulation, 54 International Economic Review 307, 312 (2013). See also Anna

GAFTA requires direct transportation between members.[38] Direct transportation applies to goods which satisfy the origin requirements and which are transported directly between GAFTA parties. However, transportation through third countries is allowed under strict conditions requiring supervision by customs authorities of the transit countries.[39] In addition, a good will lose its originating status if, subsequent to meeting the required ROO, the good undergoes further production operations outside the territories of Arab countries, other than operations related to shipment of the good such as loading or unloading.[40] As a result of this exception from the direct transportation rule, GAFTA did not need specific provisions for goods sent to exhibitions outside the GAFTA region. Nevertheless, GAFTA includes detailed rules on exhibitions i.e. originating goods which were sent to countries outside GAFTA for exhibition in third countries and which are sold to a buyer in GAFTA. These goods sent for exhibition benefit from preferential treatment at importation. If there is no exhibition exception, such goods would have to be returned to the GAFTA country where the good had initially been manufactured for re-export to the GAFTA country of the buyer to benefit from the preferential treatment.

GAFTA prohibits duty drawback.[41] In other words, GAFTA does not reimburse tariffs paid on non-originating components that are subsequently included in a final product exported to another GAFTA country. This is to encourage companies to change from imported components from non-participating GAFTA countries towards sourcing inputs from GAFTA countries. It could also be that duty drawback is prohibited so as not to favor producers who direct their products-using non-originating components- toward export over producers who direct their products to the domestic market. Producers for the domestic market are put at disadvantage. Thus, to ensure equal footing in treatment, GAFTA parties ruled in favor of this prohibition. However, the absence of duty drawback – combined with tariffs and logistical costs- may lead to an increase in the cost of final product as a result of incorporating components with no drawback. GAFTA could allow for the use of partial drawback subject to certain qualifications.[42]

---

Jerzewska, Brexit and Origin: A Case for the Wider Use of Cross- Cumulation, International Centre for Trade and Sustainable Development 1, 2 (2018), available at
<https://www.ictsd.org/sites/default/files/research/rta_exchange_-rules_of_origin-jerzewska-final.pdf> (cross-cumulation is increasingly being applied in trade agreements around the world. Cross cumulation allows the cumulation of origin between three or more countries which are not necessarily joined by a trade agreement or are joined by agreements with disparate rules of origin. For example, under the Canada-Colombia and Canada-Peru trade agreements, various materials originating in the US can be used in the production of passenger vehicles while maintaining originating status).

[38] See Rule 17, Chapter 4, GAFTA. See also General and Detailed Principles of GAFTA Rules of Origin, Doc. No. G03-19/13 (07/09)/02-t (0542)).
[39] Maria Donner Abreu, Preferential Rules of Origin in Regional Trade Agreements, WTO Staff Working Paper No. ERSD-2013-05, p. 13 (2013).
[40] See Rule 17, Chapter 4, GAFTA, *supra* note 35.
[41] See General and Detailed Principles of GAFTA Rules of Origin, art. 14, Doc. No. G03-19/13 (07/09)/02-t (0542)). Drawback can be defined as the refund or remission, in whole or in part, of a customs duty which was imposed on imported merchandise because of its importation. See Munford Page Hall and Michael S. Lee, International Trade Decisions of the Federal Circuit, 57 Am. U.L. Rev. 1145, 1158 (2008). Countries use duty drawback to attract investment and encourage exports. See also Arvind Panagariya, Input Tariffs, Duty Drawbacks, and Tariff Reforms, 32 Journal of International Economics 131(1992).
[42] For example, partial drawback can be used for particular products. In addition, a limit on the amount of refundable duties can be imposed under partial drawback. Partial drawback should be subject to time

GAFTA does not include a provision for fungible goods or fungible materials.[43] Certain accounting methods can be used to determine the different origin of input materials which are identical and interchangeable without any obligation to stock non-originating and originating physically segregated.[44] These methods are not used for finished products but only for inputs. When modifying GAFTA rules, these rules should not make such distinction regarding the application of these methods. In addition, any GAFTA modification should set pre-conditions for applying these accounting methods such as requiring a specific authorization from the customs authorities for the application of this method. The purpose of all these modifications should be to facilitate trade and ensuring that these accounting methods for fungible goods do not become unnecessarily restrictive.

In contrast with GAFTA detailed preferential ROO, GAFTA does not include provisions for determining when a product which does not meet the preferential ROO. Nevertheless, it will be considered a product of a particular Arab country for other purposes such as the application of a country-specific quota or anti-dumping duty or countervailing duty.

Besides ROO for goods, GAFTA does not establish ROO for services.[45] This is major omission considering that most economies are service-driven. The rationale for establishing ROO for goods also applies for services. However, the mechanism for determining origin is totally different. For example, while the focus of ROO for goods lies primarily in the manufacturing of the good (for example added-value test or change in tariff heading), the focus of ROO for services is on the nationality of the service provider.[46] Nationality can be determined by country of nationality for natural person and by ownership or control for companies. Of course, GAFTA can be modified, by virtue of article 36 chapter 4 of GAFTA which permits the Economic and Social Council to change ROO, to accommodate ROO for services.

**IV. Review of Customs Decisions Regarding GAFTA ROO**

Importers and exporters should have the right to request a review of decisions given by the customs authorities. GAFTA does not have specific provisions on review and appeal. In GAFTA, review and appeal of administrative actions are dealt with under the general customs law.[47]

---

limits and periodic reviews. See Sherzod Shadikhodjaev, Duty Drawback and Regional Trade Agreements: Foes or Friends? 16 Journal of International Economic Law 587, 609 (2013).

[43] Fungible goods or fungible materials means goods or materials which are interchangeable because they are of the same kind of commercial quality, possess the same technical and physical characteristics, and, once mixed, cannot be readily distinguished. See Arthur E. Appleton and Michael G. Plummer, The World Trade Organization: Legal, Economic and Political Analysis 647-648 (2007).

[44] *Id*.

[45] This is understandable since GAFTA does not cover service which reduces the benefit of the agreement. See Bernard M. Hoekman and Jamel Zarrouk Catching Up with the Competition: Trade Opportunities and Challenges for Arab Countries 297 (2000).

[46] See Olivier Cadot, Antoni Estevadeordal, Akiko Suwa-Eisenmann, and Thierry Verdier, The Origin of Goods: Rules of Origin in Regional Trade Agreements 135-138 (2006). See also Khuong-Duy Dinh, "Mode 5' Services and Some Implications for Rules of Origin, 12 Global Trade and Customs Journal 299, 301 (2017).

[47] See General and Detailed Principles of GAFTA Rules of Origin, art. 29.3, Doc. No. G03-19/13 (07/09)/02-t (0542)).

Perhaps the most important aspect of GAFTA ROO is the dispute settlement process. Any dispute relating to interpretation of ROO or origin verification is settled bilaterally or if unsuccessful through the Committee on Implementation and Execution or through the Committee of Dispute Settlement.[48] The dispute settlement process can play a crucial role especially at times when the political will of the integration is questionable. If the dispute settlement body or process is seen independent and able to ascertain its power, this will engender confidence in an integration scheme.[49] There is no public record of origin disputes among GAFTA countries that would shed light on thorny issues. The resulting body of case law can help in interpreting origin rules and remove ambiguities.

## V. The Status of ROO in GAFTA

There are several factors that affected the negotiations for ROO starting from drafting the agreement on facilitating and developing trade among Arab countries of 1981. Article 9 of the agreement provided the Economic and Social Council will decide on ROO for purposes of the agreement taking into account that the value-added for goods produced in an Arab country should be no less than 40% and this percentage would be reduced to 20% in case of industrial assembly projects in Arab countries.

In 1984, an expert committee for tariffs determined the criteria for setting the added-value on the basis of the difference between the final value of the good and value of imported raw materials or imported inputs used in the production of the good in question. Then, in 1995, the World Trade Organization (WTO) created the Agreement on Rules of Origin which provided rules for determining non-preferential treatment.[50] Nevertheless, Arab countries - during the transition period of implementing the executive program of GAFTA- were guided in part by the WTO Agreement on Rules of Origin. For example, GAFTA requires that ROO do not themselves create restrictive, distorting or disruptive effects on trade among Arab countries.[51] In addition, GATT ROO is based on a positive standard i.e. ROO should state what does confer origin rather than what does not. Negative standards are permissible either as part of a clarification of a positive standard or in individual cases where a positive determination or origin is not necessary.[52] New ROO or modifications thereof should not apply retroactively.[53] During the transitional period, Arab countries adopted decisions regarding the general application of ROO and any conditions that must be satisfied especially in case of a change of tariff category test and any exceptions for this rule.

---

[48] *Id*. art. 29.1 &2.
[49] In the North American Free Trade Area (NAFTA) integration example, panels played an important role in strengthening the free trade area. See also Myung Hoon Choo, Dispute Settlement Mechanisms of Regional Economic Arrangements and Their Effects on the World Trade Organization, 13 Temp. Int'l & Comp. L.J. 253, 254 (1999). See also Ji-Soo Yi, A Study on the Dispute Settlement Procedure for the Preferential Rules of Origin, 26 J. Arb. Stud. 3, 6-8 (2016).
[50] Article 1(1) of the WTO Agreement on Rules of Origin limits its scope to "those laws, regulations and administrative determinations of general application applied by any [WTO members] to determine the country of origin of goods provided such rules of origin are not related to contractual or autonomous trade regimes leading to the granting of tariff preferences going beyond the application of paragraph 1 of Article I of GATT 1994." Article 1(2) of the WTO Agreement on Rules of Origin then states that the ROOs referred to in Article 1(1) include "all rules of origin used in non-preferential commercial policy instruments, such as in the application of . . . most-favored-nation treatment under Articles I." See World Trade Organization, Agreement on Rules of Origin, 1868 U.N.T.S. 397 (1994). See also Zviad V. Guruli, What is the Best Forum for Promoting Trade Facilitation? 21 Penn St. Int'l L. Rev. 157, 163 (2002).
[51] See GAFTA, Chapter 4, rule 10.
[52] The use of "negative standards if they clarify a positive standard" is vague as to have had small impact. *Id.* rule 12.

In 1996, Arab countries entered into new round of negotiations for adopting ROO. This came in response to an earlier decision taken by the Economic and Social Council of LAS which clarified that according to paragraph 1 of the Economic and Social Council of LAS No. 1249 that Arab countries should submit its own ROO and determine goods that should receive priority for determining its origin. Over the years, the LAS technical committee for ROO have held many meetings independently and sometimes in conjunctions with the meetings of the Economic and Social Council for purpose of discussing detailed rules of origin. For example, on Sep. 6, 2007, the Economic and Social Council approved the general rules for ROO for the list of goods agreed upon.

By examining the different reports and recommendations issued or made by the LAS technical committee for ROO as well as the extraordinary rounds for the Economic and Social Council, Arab countries of GAFTA are in agreement for some ROO. However, there remain ongoing negotiations among parties for adopting ROO for some other goods. There are several economic factors that make these goods sensitive from the perspective of some Arab countries: number of economic industries, production volume, level of investment, economic contribution to economy, export value, and use of large number of labor.[54] Many GAFTA Member States face pressures from particular interests to delay or avoid the implementation of some of ROO. It is noted that only nine of eighteen Arab countries submitted statistics concerning sensitive goods.[55] Moreover, those nine countries submitted these statistics at a later date after several calls and recommendations of LAS bodies. Even the statistics submitted by these countries were absolute not relative/comparative i.e. not of total economic industries in the country in question, export value, labor. This is contributed to delayed implementation of ROO.

A Saudi-Moroccan working group was formed to develop a proposal for those goods whose ROO are disagreed upon. On Oct. 27, 2011, the outcomes of the working group were discussed in a meeting of the technical committee of ROO which approved some of ROO for a number of goods. The technical committee on rules of origin recommended in its 21 meeting during the period April 8-11, 2012 continuing discussion of ROO for those goods which Arab countries have not agreed on their origin determination. It is noted that there is a gap in the timely approval from the concerned bodies in LAS. For example, there was a delay in adopting recommendations of the technical committee on rules of origin as the adoption of the recommendations did not occur until the Economic Summit No. 1924 of 2013 in the ordinary meeting No. 91 held during the period Feb. 11-12, 2013.

The technical committee for rules of origin held meeting no. 24 on May 9-7, 2013. In addition, the General Secretariat of LAS issued a circular of the report and recommendations of the Doha Summit no. 582 held on March 24-26, 2013. The circular stated in part that 80% or more of Arab countries are principally in consensus regarding ROO.[56] The circular also encouraged those remaining Arab countries to improve their negotiating positions before the end of 2013. The technical committee on rules of origin would continue to discuss those ROO that are disagreed upon and present the outcome of the discussion the Economic and Social Council in its next meeting. Some Arab countries (Tunisia, Sudan, Algeria, Egypt, Morocco, and Yemen) made reservations concerning adopting a majority voting when approving ROO. For those countries, ROO are related to economies of countries and the status of industries in these countries –protecting Arab industries and preventing non-originating imports from enjoying preferential treatment- and thus decisions should be based on consensus. Therefore, it is obvious that there is disagreement among Arab countries on the decision mechanism for adopting ROO.

---

[53] *Id*. rule 14.
[54] See Explanatory Note, Extraordinary Meeting of the Economic and Social Council S (15/06)/02-M(0271), pp. 41-46 (June 14-16, 2015).
[55] *Id*.

In light of the above, the Economic and Social Council issued decision instructing the General Secretariat of LAS to prepare a list of those goods where there is disagreement and distribute the list to countries by the end of February 2016.[57] Also, the Economic and Social Council requested Arab countries members of GAFTA who have sensitive goods on the list to provide the General Secretariat of LAS of these goods and any supporting information. Nevertheless, in the extraordinary meeting of the Economic and Social Council held during May 20-22, 2015, only four countries provided an initial list of important goods.[58] The delayed submission by other Arab countries is considered an obstacle in concluding negotiations over ROO.[59]

-In the extraordinary meeting of the Economic and Social Council held during Sep. 3-4, 2016, participants were informed of the classification made by the General Secretariat of LAS for lists of goods whose ROO are not agreed upon. These lists are:

  List (1): Goods not included in GAFTA members' lists which include 37 rules

  List (2): Goods of relative importance for industry coming from one country which include 24 rules.

  List (3): Goods of relative importance for industry coming from two countries which include 9 rules.

  List (4): Goods of relative importance for industry coming from three countries which include 8 rules.

List (5): Goods of relative importance for industry coming from four countries or more which include 47 rules.

Those participants in the meeting agreed to apply the general rule of origin (40% value-added for goods included in lists 1, 2, and 3). Participants agreed to hold an extraordinary meeting to discuss goods on lists 4 and 5. Arab countries were supposed

---

[56] *Id*. p.6.
[57] See League of Arab States, Economic and Social Council, Decision no. 1984 (Feb. 13, 2015).
[58] These four countries are: Egypt, Palestine, Morocco, and Yemen.
[59] Indeed, this is what forced those attending the extraordinary meeting of the Economic and Social Council to recommend to all Arab countries to submit the requested information (number of industry establishments in each country, labor, production volume, investment level, inter-Arab import and export) within 30 days of the meeting. Then, the General Secretariat of LAS would refine the list of goods and send back to countries within 30 days.

to provide the General Secretariat of LAS with ROO to facilitate negotiations over lists 4 and 5.

In the extraordinary meeting of the Economic and Social Council held in the period Jan. 5-6, 2017, participants discussed lists of 4 and 5 which include 55 goods of relative importance for three Arab countries or more. Participants agreed on two lists: List (a) where there is consensus over its rule of origin and includes 12 good items and list (b) where there is no consensus over its rule of origin and which includes 43 good items. Participants recommended approval of list (a). As for list (b), it would be subject to further discussion during the next meeting of the ordinary meeting of the Economic and Social Council. In the meantime, list (b) will be subject to the general rule of origin of 40% value-added taking into account interests of least developed Arab countries. It is noted that six countries (Tunisia, Sudan, Algeria, Egypt, Morocco, and Libya) made reservations regarding this recommendation claiming that it is contrary to earlier decisions of the Economic and Social Council to agree on all ROO as part of a single package. Those six countries also disagreed over the application of the general rule of origin of 40% value-added regarded list (b) goods. Those six countries will take several measures to protect their local industries and their goods until their ROO are agreed upon especially that those six countries offered greater flexibility to finalize ROO for these goods during the meeting.

The Economic and Social Council of LAS held an extraordinary meeting on June 14-16, 2017 to examine several proposals for these 43 good items subject to disagreement among Arab countries. In the meeting, participants classified these 43 good items into three lists and recommended: approval of ROO for certain goods (List 1) on the basis of change of tariff heading. The total number of these items on the list is 16 and they vary from pharmaceutical goods, polyester, shoes, steel, furnaces and oven, aluminum products, electrical cables, to metallic furniture. ROO for these items look for downstream input vs. processed product which ought to fall into a different class or kind when compared to the downstream input. To put it differently, ROO are met when all downstream input and upstream product falls into two different classes.

It should be noted that Bahrain, Saudi Arabic, Oman, and Qatar made reservations over this list and proposed that the approval of ROO for List 1 items should be postponed until ROO for the remaining two lists of items (List 2 and 3) are agreed upon all together. It is more practical to get the implementation of ROO on List 1 items forward regardless of the approval on the remaining list of items. Otherwise, the implementation of ROO for the agreed upon items on List 1 can be delayed and held hostage for the outcome of negotiations on Lists 2 and 3 items which could take months if not years to conclude. In addition, unlike World Trade Organization (WTO), there is no legal requirement of "single undertaking" under GAFTA.[60] In other words, Arab countries do not have the obligation to accept all the negotiations results or nothing.

List 2 includes items under negotiations revolve around the appropriate value-added percentage. The total number of items on List 2 is 19 mostly of manufactured goods such as pesticides, plastic goods, air conditionings, refrigerators, TVs, radios, cables, and vehicles. Some Arab countries suggested adopting two options for foreign imported inputs: 45% or 55% as a percentage of the final product price. On the other hand, some Arab countries –Jordan, UAE, Bahrain, Saudi Arabia, Iraq, Oman, and Qatar- made reservations concerning the foreign input percentage and suggested that the percentage should be 60% instead of 55%. The different propositions made by Arab countries can be attributed to diverged economic interests of these countries. There are assembly operations/plants in Arab countries that employ large number of workers and provide job opportunities. In the short and medium run, the more foreign input in the value-added test the better in meeting the proposed ROO for items on List 2. The reasoning for such a suggestion (55% or 60%) is to easily meet ROO in GAFTA without it would be difficult to claim Arab country origin. Moreover, it would be better to maintain the level investment in these manufacturing areas and level of employment. In the long run, for purposes of development, the local content should be increased in return for decreasing the percentage of foreign input. This will strengthen local industry in GAFTA countries with focus on innovation, Research & Development, and intellectual property.

---

[60] Single undertaking is known also as the principle of globality or single roof policy. See Anu Bradford, When the WTO Works, and How It Fails, 51 Va. J. Int'l L. 1, 54 (2010).

List 3 includes items not agreed upon and still under negotiations. The total number of items on List 3 is 8 and includes goods such as yogurt and cream, cereals, meat and fisheries, sausage, sugar cane, fruits juice, natural and sparkling water, cloths and cloths accessories. Arab countries have different negotiation positions regarding the items on List 3.[61] For example, for animal products (yogurt and cream) positions of Arab countries can be grouped into two groups: 1) those agricultural Arab countries such as Algeria, Egypt, Jordan, Morocco, Sudan, Syria, and Tunisia who favor the wholly-produced rule. Regardless of the domestic interests in these countries, the wholly-produced rule is logical and in line with GAFTA ROO (Chapter 4, Rule 7/D of GAFTA); 2) those Arab countries who are not agriculture by nature and have local industry that depends upon inputs from other countries and thus they go for the 50% value-added rule. The application of cumulation rule in this area is helpful to alleviate

---

[61] For the sausage item, most of Arab countries proposed 50% value-added content. On the other hand, there are other countries that proposed to adopt the wholly-produced rule. The value-added criterion is more logical than the wholly-produced rule although sausage is taken from animals but it needs further processing and thus value-added. For sugar and sugar cane, positions of Arab countries vary between adopting specific manufacturing process starting from refinement stage, changes of tariff heading, 60% value-added, and wholly-produced rule. Eight Arab countries are calling for adopting specific manufacturing process starting from refinement stage. It is preferable to adopt the specific manufacturing process starting from refinement stage because this is the most significant and valuable step in producing sugar. Change in tariff heading may not capture the actual transformation process of making sugar. For fruit juice, nine Arab countries are calling for a 60% content while eight Arab countries are suggesting adopting combined criteria of change in tariff heading and 30% content. Those countries calling for the 60% content are not agricultural countries and do not produce fruits rather they import them. Those countries that call for combined criteria of change in tariff heading and 30% content are agricultural countries depending on local produce of fruits and thus percentage of foreign imported input is low. Applying the cumulation rule in this area is helpful. For waters, including mineral waters and aerated waters, containing added sugar or other sweetening matter or flavored, and other non-alcoholic beverages, not including fruit or vegetable juices of heading 2009, Arab countries are in disagreement as to the appropriate rule of origin. Some Arab countries are calling for change of tariff category test whereby all parts used in manufacturing be classified under a different tariff category than the category in which the final product would be classified. As Alternative rule of origin, the ex factory price of any non-originating materials should not exceed 50%. For articles of apparel and clothing accessories, knitted or crocheted (chapter 61), some Arab countries (such as Tunisia a textile-producing country) are calling for manufacturing processing starting from thread i.e. spinning stage, while other countries (such as Lebanon) are calling for manufacturing processing starting from fabric.

differences between these two groups and can enhance inter-trade among Arab countries.

Examples of gaps between decisions and implementation include observations made among numerous exporters that despite compliance with ROO under GAFTA, companies are not granted preferential treatment and are obliged to pay tariffs.[62] Such decisions are often the result of customs officials' lack of knowledge about the different origin requirements. Moreover, for GAFTA, the zero tariff rate has been applied for the past few years, but only for goods whose rules of origin are agreed upon.

**VI. GAFTA Certificate of Origin**

While certificates of origin are necessary, they also generate cost and delays for businesses and may be a source of risk and uncertainty.[63] According to GAFTA, the certificate of origin is prepared by the exporter or his representative.[64] However, GAFTA should allow also the producer of the goods to prepare the certificate since the exporter may reasonably rely on the written representation of the producer that the good satisfies the particular rule of origin upon which the claim of preferential treatment under GAFTA is founded. After completion of the certificate by the exporter, it shall be issued and certified by designated authorities in Arab countries members of GAFTA. Issuers of certificates of origin under GAFTA can be categorized into four categories:[65]

1) Chambers of commerce (Jordan, Tunisia, Sudan, Syria, Iraq, Palestine, Qatar, Lebanon, Libya, and Yemen).
2) Ministries of Economy/Trade/Industry (Kuwait, UAE, and Saudi Arabia).
3) Customs authorities (Bahrain and Morocco).
4) Import and export authorities (Egypt).

In general, chambers of commerce are the leading entities for issuing and certifying certificates of origin under GAFTA. Other governmental entities play less important role. There should a standardized authority in all Arab countries that oversee certificates of origin i.e. chambers of commerce.

The certificate itself is a relatively simple and straightforward document. It contains information relating to the goods in question, producer, exporter, importer, date of manufacturing, final value of the goods, and official stamps. The exporter must simply follow the instructions and complete the form. However, the most time-consuming part of the certificate of origin requirement is determining how goods qualify for preferential tariff treatment under GAFTA. Once the exporter has made that determination, the exporter can easily insert the information on the certificate by simply following the instructions provided. Thus, the additional burden imposed by GAFTA certificate of origin, as a stand-alone document- is minimal.

---

[62] See International Trade centre, Making Regional Integration Work – Company Perspectives on Non-Tariff Measures in Arab States 27 (2015).
[63] See WTO, WTO Hosts Forum to Exchange Views on Trade Challenges from Certificates of Origin (April 18, 2018).
[64] See General and Detailed Principles of GAFTA Rules of Origin, art. 16, Doc. No. G03-19/13 (07/09)/02-t (0542)).
[65] See GAFTA, Part 4, Annex 1, Executive Program for Implementing GAFTA.

The exporter must complete the certificate in the Arabic language or can be translated to another language whenever it is necessary.[66] A number of countries require the GAFTA certificate of origin and enclosed bills to be 100% in Arabic.[67] This is a strict interpretation of GAFTA rules. The purpose of strictness is evidently to protect the interests of importing country and prevent against concealing fraud. However, it makes common sense to allow some items to be in English especially if these items are commonly used. Trade could not proceed if GAFTA rules are interpreted strictly as there must be a level of good faith and honesty.

GAFTA regulations focus on paper format for proof origin.[68] This is understandable since GAFTA approved in 1990's when the use of Internet was not common. However, with the advance of technology electronic format should be acceptable to facilitate trade.

The certificate may cover a single importation of goods or multiple importations of goods (blanket certificate).[69] The certificate of origin is valid for a six-month period.[70] The certificate shall be submitted to the customs authority of the importing country within the validity period. However, even though the certificate is submitted after the period specified, the certificate can be accepted in case the failure to observe such a period because of *force majeure* or other circumstances accepted by the importing country. For purposes of flexibility and facilitating inter-trade, the validity period for certificate of origin can be expanded to twelve months for instance. For the same token, exporters may need to include only a copy of the certificate with other documentation submitted to customs authorities. There is no need for certificates with original signatures as a certified copy suffices.

The manufacturing record and all other pertinent records which substantiate the claim for preference based upon GAFTA origin must be retained by the exporter for three years.[71] Thus, GAFTA provides for a relatively long record-keeping period. Throughout this period, the substantiating records can be available for an Arab country customs services - seeking to verify origin- upon demand. Similarly, the importer, in reliance on the certificate of origin, must repeat the claim for preferential tariff treatment under GAFTA as part of his/her entry documentation.[72] There is no requirement that the importer must retain his/her records, including the certificate of origin and all other relevant documentation, for a period of time. Retaining such documentation can help to avoid denial of preferential tariff treatment or the imposition of penalties.

---

[66] See General and Detailed Principles of GAFTA Rules of Origin, art. 16, Doc. No. G03-19/13 (07/09)/02-t (0542)).
[67] See International Trade centre, *supra* note 59, at 28.
[68] Electronic application and issuance of preferential certificate of origin are even less implemented. Electronic exchange of certificates of origin is the least implemented measure having been implemented on a limited basis by less than 10% of the economies in the region. Joint United Nations Regional Commissions, Trade Facilitation and Paperless Trade Implementation Survey 2015: Middle East and North Africa Report 23 (2015).
[69] See General and Detailed Principles of GAFTA Rules of Origin, art. 22, Doc. No. G03-19/13 (07/09)/02-t (0542)).
[70] *Id*. art. 20. The validity period means that the duration allowed for finalizing the importation of goods under this certificate from the date of issuance in the exporting country.
[71] *Id*. art. 25.
[72] *Id*. art. 21.

Custom authorities in Arab countries initially afford duty free treatment to shipments of goods accompanied by appropriate certificates of origin. Denial of benefits may occur only after a verification inquiry which determines that GAFTA treatment was unwarranted.[73] The regulations provide procedures for verifying the accuracy of certificates of origin, including but not limited to checking documents. This could include checking for signatures, stamps, and accuracy of the content. Verification could also mean visiting a manufacturer or exporter in the exporting Arab country.

The customs authority of the Arab importing country request from the relevant authority of the Arab exporting country to undertakes verification.[74] This requires mutual administrative cooperation between the authorities in the importing and exporting countries.[75] Authorities in the exporting country then conducts an investigation and checks the accuracy of a proof of origin or the invoice declaration or any other related documents. The Customs authority of the exporting country presents the findings and results of the origin determination to the customs authority of the importing country.[76] Based on current provisions of GAFTA, the customs authority of exporting country plays a larger role in verification. There is no direct interaction between exporters in the exporting country and the customs authority of the importing country.

The importing Arab country has the strongest interest in determining whether it has been deprived of revenue through improper claims for preferential tariff treatment. In this instance, inter-country verification missions could prove helpful. GAFTA does not provide for governmental authorities of one Arab country to make verification into another Arab country's territory. If permitted that may lead to sensitive jurisdictional issues regardless of the fact that such verifications are important. However, as a consequence of this anticipated sensitivity, GAFTA could have provided for a notification and consent process for verifications, along with representation by counsel and the presence of observers. In other words, through notification and consent, customs authorities of one Arab country may not undertake a verification mission if consent is refused. In the absence of consent or upon finding that the required supporting documentation is missing, the customs authorities presumably will deny GAFTA tariff treatment to the goods in question (can be called adverse conclusion).[77]

To facilitate trade under GAFTA, Arab custom authorities can indentify certain goods as priorities for verifications. These goods may include certain agricultural goods, certain electrical products, and apparels. These goods, and possibly others, can be singled out for deeper verification presumably because they could represent large volumes of imports or politically sensitive. It is not incidental that many of these goods are among those subject to complex rules of origin.

---

[73] *Id*. arts. 16.5, 28.
[74] *Id*. art. 28.2.
[75] In some countries (such as Kuwait, Egypt, and Morocco), customs authorities in the importing countries tend to experience some difficulties in securing the necessary cooperation from the responsible authorities in the exporting countries. See World Customs Organization, Origin Irregularity Typology Study, 7-8 (July 2013).
[76] See General and Detailed Principles of GAFTA Rules of Origin, art. 28.5, Doc. No. G03-19/13 (07/09)/02-t (0542)).
[77] *Id*. art. 28.6

In a step taken to facilitate inter-trade among GAFTA countries, it is acceptable to present a commercial invoice –as long as it is issued in an Arab country- as proof of origin.[78] This step presents another option for traders to proof origin thus reducing non-compliance and ease administrative procedures and costs (no need to fill in certificate of origin and stamp it from designated authorities). Exporters can simply insert the required preferential origin declaration on a commercial invoice.[79] However, details on the use of commercial invoice as proof of preferential treatment must be spelled out. Details would have to determine the list of approved exporters who make frequent shipments and are authorized to make invoice declarations without the direct involvement of any issuing authority.

Because of the legitimate need of exporters and importers to know whether GAFTA preferential tariff rate or the higher most-favored nation tariff rate will apply to their products, GAFTA regulations should allow written advance rulings. Advance ruling allow traders to obtain an origin ruling prior to the importation of a product. Hence, advance ruling saves time and energy.[80] Advance rulings can be issued within say 90 days of a request. Penalties are imposed against a requesting party that has omitted or misrepresented material facts in its request for a ruling.

There is a form of leniency for trivial discrepancies in the information contained in the certificate of origin and accompanied documents. In case of no significant errors, GAFTA does not nullify the already issued certificate of origin or requires the issuance of a new certificate of origin.[81] Rather, the certificate of origin could be changed. In practice, however, forms are sometimes rejected because of minor mistakes such as typos.[82] It seems that in cases of doubt, the safest course customs official to take is to reject the forms. A distinction must be made between minor and major variations or discrepancies and whether it is material enough to warrant rejection. For example, if there variation relates to unimportant typos then the certificate of origin as well as accompanied documents should be accepted. On the other hand, if discrepancies relate to the description of goods then a judgment can be made to reject the documents. At any rate, rejection of certificate of origin and any accompanying documents should not be made for petty bureaucratic reasons.

Companies are responsible for other forms of discrepancies between preferential and regular tariff rates where origin is denied, and authorities will require the importer to pay the difference between GAFTA duties and the normal most favored nation duty. In addition, customs authorities can collect these fines. The question that arises is whether fines should be collected retroactively or from the date of denial. In the optimal scenario, it is better if fines are collected from the date of denial especially for *bona fide* companies.

---

[78] See The Economic and Social Council, Decision No. 1960 (Sep. 12, 2013).

[79] Commercial invoice is usually used to determine the actual transaction value of a particular shipment of goods.

[80] See Tim Tatsuji Shimazaki, North American Free Trade Agreement: Rules of Origin--Free Trade or Trade Barrier? 25 W. St. U. L. Rev. 1, 24-25 (1997) (NAFTA contains a provision requiring on-demand advance rulings). See also Bashar Malkawi, Rules of Origin under U.S. Trade Agreements with Arab Countries: Are they Helping and Hindering Free Trade? 10 Journal of International Trade Law and Policy 29, 34 (2011).

[81] The discovery of minor discrepancies between the information recorded in the certificate of origin and those made in the documents submitted to the customs authorities of the importing country for the purpose of carrying out the procedures for importing the products shall not *ipso facto* invalidate the certificate of origin if it does in fact correspond to the products submitted. See General and Detailed Principles of GAFTA Rules of Origin, art. 26, Doc. No. G03-19/13 (07/09)/02-t (0542)).

[82] See International Trade centre, *supra* note 59, at 28.

There are times when certificate of origin is not required. Generally, importers do not need certificates when the value of an imported good does not exceed US $500 for small parcels or US $1000 for goods that constitute part of the personal effects to a person.[83] Of course, to benefit from this exemption, importation should not part of a series of importations since this may reasonably be considered to have been arranged for the purpose of avoiding the certificate requirement.

A delay in issuing certificate of origin is a cause of concern for exporters. Because certificates are issued much later than expected or promised, products may be held at the production site, the home border or the port of destination, resulting in delays in shipment and demurrage fees.[84] Delays in obtaining certificates of origin are aggravated by the paperwork that accompanies certification requests with unnecessary duplication and at times differing demands depending on the day or the official in charge. Delays are often reported when more than one institution is involved in the process due to limited coordination among them.[85] It seems that procedures related to the certificate of origin remain opaque and defined in an *ad hoc* way.

To encourage regional trade among GAFTA countries, GAFTA regulations need to be re-examined to allow for exporter's self-determination of origin as this can help in facilitating prompt clearance.[86] Of course, there is always the risk that such self-determination will be accepted by customs authorities in the Arab country in question. However, such self-determination can balance duties of customs authorities and importers for the sole purpose of facilitating trade.

**VII. Impact of Rules of Origin on Trade among GAFTA Countries**

The experience under GAFTA suggests that its goals are being met steadily. However, intra-trade among GAFTA countries is still hovering around 10 percent of the total trade of Arab Countries.[87]

In 2016, the share of Arab exports to other countries was around 4.7% of total world exports, whereas its portion of imports did not surpass 4.6% of total world imports. The figures of table (1) reveal the extent of Arab countries contribution to world trade

---

[83] See General and Detailed Principles of GAFTA Rules of Origin, art. 23, Doc. No. G03-19/13 (07/09)/02-t (0542)).
[84] See International Trade centre, *supra* note 59, at 29.
[85] *Id*.
[86] See Patricio Díaz Gavier and Luc Verhaeghe, The EU–Korea Free Trade Agreement: Origin Declaration and Approved Exporter Status, 7 Global Trade and Customs Journal 315, 310 (2012) (The EU-Korea free trade agreement is a new generation agreement. The system of origin certification by a competent authority is replaced by a system of self-certification while in previous free trade agreements both systems coexisted. Under this free trade agreement, consignments of EUR 6000 or more can only be granted preferential treatment where the origin declaration is made out by an approved exporter).
[87] See Imed Limam and Adil Abdalla, Inter-Arab Trade and The Potential Success of AFTA, Arab Planning Institute, API/WPS 9806, Kuwait, p.1, available at <http://www.arab-api.org/images/publication/pdfs/305/305_wps9806.pdf> (last visited Oct. 10, 2018). See also Hassan Al-Atrash and Tarek Yousef, Intra-Arab trade: Is too little? IMF Working Paper n. WP/00/10, (2000), p. 18.

during the years 2012-2016. It is noticed that the contribution of Arab countries is insignificant.

**Table (1): Evolution of Total Arab Foreign Trade during the Period 2012-2016.**

|  | Value (US$ billion) | | | | | (%) of total world exports (imports) | | | | | Annual change rate (%) |
|---|---|---|---|---|---|---|---|---|---|---|---|
|  | 2012 | 2013 | 2014 | 2015 | 2016 | 2012 | 2013 | 2014 | 2015 | 2016 | 2012-2015 |
| Arab exports to other countries | 1322 | 1311 | 1244 | 857 | 769 | 7.2 | 7.0 | 6.6 | 5.2 | 4.7 | 1.4 |
| Arab imports from other countries | 816 | 858 | 901 | 851 | 796 | 4.4 | 4.5 | 4.7 | 5.1 | 4.6 | 13.5- |

**Source**: **Arab Monetary Fund, Joint Arab Economic Report, 2017, pp. 143 and 367.**

On the other side, intra-trade among Arab countries as percentage of their total exports and imports was respectively around 12% and 14% in 2016. The ratio of intra-Arab exports has been increasing steadily during the period 2012-2016, while the percentage of intra-Arab imports has remained relatively constant over the reference period. Some quantitative studies concluded that overall intra-Arab trade should be about 10-15 percent higher than what is observed.[88]

**Table (2): Evolution of Intra Arab Trade during the Period 2012-2016.**

|  | Value (US$ billion) | | | | | (%) of total Arab exports (imports) | | | | | Annual change rate (%) |
|---|---|---|---|---|---|---|---|---|---|---|---|
|  | 2012 | 2013 | 2014 | 2015 | 2016 | 2012 | 2013 | 2014 | 2015 | 2016 | 2012-2015 |
| Intra-Arab trade exports | 112 | 117 | 121 | 107 | 96 | 8.4 | 9 | 10 | 12.4 | 12 | 1.4- |
| Intra-Arab trade | 111 | 121 | 122 | 115 | 110 | 13.6 | 14 | 13.5 | 13.5 | 14 | 1.2 |

---

[88]See Hassan Al-Atrash and Tarek Yousef, Intra-Arab trade: Is too little? IMF Working Paper n. WP/00/10, (2000), p. 18.

| imports | | | | | | | | | | |

**Source: The figures in columns 1 and 2 were derived from Arab Monetary Fund, Joint Arab Economic Report, 2017, p. 370, whereas figures in column 3 were calculated by the authors.**

The top exporting countries are Saudi Arabia, UAE, and Egypt capturing 64.6% of total intra-regional trade in 2016.[89] The top importing countries are UAE, Saudi Arabia, and Oman with 49.8% of total intra-regional trade.[90] Roughly half of Saudi Arabia exports to the region are oil-related and much of the remainder is related to downstream hydrocarbon industries such as petrochemicals, plastics, rubber and chemicals. Saudi Arabia is also an important regional source of manufactured goods and food. The UAE is an important player to intra-regional trade as non-hydrocarbons account for major portion of UAE exports. However, around 38% of exports from the UAE are re-exports mainly from countries outside GAFTA. This means that these exports only passing through the UAE thus making it difficult to measure genuine inter-regional trade.[91]

It might be difficult to draw conclusions as how much GAFTA ROO itself contributed to volume of trade. Many factors play a role in explaining the weakness of intra-Arab trade. These factors range from mere economic factors, such as high level of tariffs and nontariff barriers, trade impediments in some Arab countries, similarity of production structure and traded goods, and lack of adequate transportation infrastructure compounded by distance.[92]

Restrictive ROO is considered as one of the major obstacles that might impede the growth of regional trade and slow down the process of economic integration by making the costs of compliance higher than the perceived worth of the underlying preference margins.[93] The empirical data on the impact of ROO on trade under GAFTA is inadequate. Thus, it is difficult to measure the effects of ROO on trade. Utilization rate is a clear sign of the efficacy of trade preferences.[94] Poor use rates of trade preferences can be explained – among various reasons- by complex ROO.[95] The different committees responsible for GAFTA should develop and make available empirical studies on pre- and post-GAFTA utilization and the impact of ROO on trade among partner countries. This would highlight the contribution of ROO to increase or decrease volume of trade under GAFTA and how ROO can be modified to facilitate trade.

---

[89] Figures were calculated by authors according the data included in Joint Arab Economic Report, Arab Monetary Fund , 2017, p. 370

[90] *Id*.

[91] See Arab News, Greater Arab free trade: Higher hydrocarbon prices boost surplus, May 13, 2013, available <http: http://www.arabnews.com/news/451429> (last visited Sep. 9, 2018).

[92] See Economic and Social Commission for Western Asia, Assessing Arab Economic Integration: Towards the Arab Customs Union, E/ESCWA/EDID/2015/4, 16-17 (2015).

[93] See Dylan Geraets, Colleen Carroll, and Arnoud R. Willems, Reconciling Rules of Origin and Global Value Chains: The Case for Reform, 18.2 Journal of International Economic Law 287, 293-294 (2015).

[94] The utilization rate of a given preferential trade agreement is calculated as the value of imports receiving preferential treatment divided by the value of imports that are eligible for preferences. See Alexander Keck and Andreas Lendle, New Evidence on Preference Utilization, WTO Working Paper ERSD-2012-12, p. 6 (2012).

[95] The existing literature and empirical studies have established that ROOs can deter preferential trade. For example, the average utilization ratio of the North American Free Trade Agreement (NAFTA) was around 64% in 2000 and in the case of the ASEAN FTA below 10% was utilized in 2002. See Jisoo Yi, Rules of Origin and the Use of Free Trade Agreements: a Literature Review, 9 World Customs Journal 43, 50 (2015). See also Kazunobu Hayakawa, Hansung Kim and Hyun-Hoon Lee, Determinants on Utilization of the Korea–ASEAN Free Trade Agreement: Margin Effect, Scale Effect, and ROO Effect,

**VIII. Conclusions**

GAFTA uses several ROO -such as change in tariff category, value-added, and certain process requirements - to determine origin. These ROO has been confusing to traders, producers, and even lawyers. No single rule used is better than the other. Each origin determination under GAFTA has its own set of advantages and disadvantages. It is noted that there are elements of unpredictability and restrictiveness in each method thus limiting the benefits of tariff reduction or elimination. For example, GAFTA requires for many goods that *all* parts used in manufacturing a good be imported and classified under a different tariff category than the category in which the final good would be classified. In order to facilitate trade, GAFTA "change of tariff category" rule should be relaxed. Also, for some goods, GAFTA requires a change in chapter category at the two-digit level from imported materials to final product (restrictive). At the other end, for some goods, GAFTA requires a change in tariff heading at the four-digit level (less restrictive). For some textiles and apparels, origin is conferred when a product undergoes the required manufacturing process or when the thread or yarn is product of an Arab country.

Some companies or enterprises in Arab countries may decide to export under the normal MFN tariffs (which can be relatively low) thus foregoing GAFTA benefits. These companies or enterprises – especially small and medium-size companies- are doing so because they believe that their costs of complying with GAFTA requirements, as reflected by change in tariff category, or value content calculations, the anticipated administrative costs of preparing GAFTA documentation such as certificates of origin, and subsequent customs verification, are greater than the import duties that would be saved. These costs may require accounting and customs specialists who small and medium-size companies do not have. This may discourage new exporters and importers from participating in the anticipated benefits of free trade under GAFTA.

Strict rules of origin limit sourcing options. There is a need to reform GAFTA ROO in several aspects. Rules of origin should as far as possible be based on a sector-by-sector rather than a product-by-product basis or line-by-line. Where the change in tariff category or specific process is used, ROO must require compliance with simple operations; where if a value-added is adopted, the percentage level should reflect the limited production capacity in some Arab countries. In addition, the use of multiple specific criteria for one product should be limited as this may lead to complexity and

---

13.3 World Trade Review 499, 507-508 (2014). See also Paul Brenton and Takako Ikezuki, The Initial and Potential Impact of Preferential Access to the U.S. Market under the African Growth and Opportunity Act, World Bank Policy Research Working Paper No. 3262, p. 21 (2004) (ROOs have a strong impact on preference utilization, particularly in apparel. Loose ROOs for LDC apparel products have had a substantial positive effect on AGOA utilization). See also Paul Brenton and Miriam Manchin, Making EU Trade Agreements Work: The Role of Rules of Origin, 26.5 *The World Economy* 755, 761 (2003) (restrictive ROOs hamper utilization of unilateral EU preferences and association agreements).

confusion. Under "net cost" rule of origin, details should be provided for allocation of other costs such as labor and overhead items. Furthermore, GAFTA should provide an option for exporters to choose the most favorable method between "net cost" and "final value of product". Full cumulation should be adopted for greater flexibility. Certain goods should be exempted from meeting the required ROO i.e. imported goods that have no tariffs at all or trivial tariff is imposed.

A review should also be undertaken of the scope for the introduction of origin certification for known reliable exporters which would ease the burden on firms in proving compliance with the requirements of the rules of origin. There must be simplification of certificates and forms of origin. Currently, there are several entities that can issue certificate of origin. Therefore, there should a standardized authority in all Arab countries that oversee certificates of origin i.e. chambers of commerce. Arabic is the language of the certificate. However, use of common technical terms in other languages such as English should be accepted. There is a need to eliminate certification for certain products shipped by less-developed Arab countries. Certificate of origin and forms should not be rejected because of minor mistakes such as typos or variations. A distinction must be made between minor and major variations or discrepancies and whether it is material enough to warrant rejection.

Institutions should adopt a tracking system for requests, complaints, and other correspondence. This would allow companies to follow the handling of their requests and shed light on what many companies perceive as cumbersome administrative procedures. Such a tracking system would also provide valuable statistics for institutions, enabling them to better monitor the quality and speed of handling requests and inquiries and to take corrective action where needed.

GAFTA regulations should allow for written advance rulings. Advance ruling allow traders to obtain an origin ruling prior to the importation of a product. A formal dispute settlement mechanism must be developed to hear ROO disputes and make it operational. The resulting body of case law can help in interpreting origin rules and cement trade liberalization.